# Sublayers Editing of Covalent MAX Phase for Nanolaminated Early Transition Metal Compounds

Ziqian Li [1,2], Ke Chen [1,3]*, Xudong Wang [1,2], Kan Luo [1], Lei Lei [1,3], Mian Li [1,3], Kun Liang [1,3], Degao Wang [1,3], Shiyu Du [1], Zhifang Chai [1,3], Qing Huang [1,3]*

[1] Zhejiang Key Laboratory of Data-Driven High-Safety Energy Materials and Applications, Ningbo Key Laboratory of Special Energy Materials and Chemistry, Ningbo Institute of Materials Technology and Engineering, Chinese Academy of Sciences, Ningbo 315201, China.

[2] University of Chinese Academy of Sciences, Beijing 100049, China.

[3] Qianwan Institute of CNiTECH, Ningbo 315336, China.

*Correspondence to:
Prof. Ke Chen, E-mail: chenke@nimte.ac.cn
Prof. Qing Huang, E-mail: huangqing@nimte.ac.cn

**Abstract:** Two-dimensional transition metal carbides and nitrides (MXenes) have gained popularity in fields such as energy storage, catalysis, and electromagnetic interference due to their diverse elemental compositions and variable surface terminations (T). Generally, the synthesis of MXene materials involves etching the weak M-A metallic bonds in the ternary layered transition metal carbides and nitrides (MAX phase) using HF acid or Lewis acid molten salts, while the strong M-X covalent bonds preserve the two-dimensional framework structure of MXenes. On the other hand, the MAX phase material family also includes a significant class of members where the A site is occupied by non-metal main group elements (such as sulfur and phosphorus), in which both M-A and M-X are covalent bond-type sublayers. The aforementioned etching methods cannot be used to synthesize MXene materials from these parent phases. In this work, we discovered that the covalent bond-type M-A and M-X

**sublayers exhibit different reactivity with some inorganic materials in a high-temperature molten state. By utilizing this difference in reactivity, we can structurally modify these covalent sublayers, allowing for the substitution of elements at the X site (from B to Se, S, P, C) and converting non-metal A site atoms in non-*van der Waals* (non-*vdW*) MAX phases into surface atoms in *vdW* layered materials. This results in a family of early transition metal Xide chalcogenides (TMXCs) that exhibit lattice characteristics of both MXenes and transition metal chalcogenides. Using electron-donor chemical scissors, these TMXC layered materials can be further exfoliated into monolayer nanosheets. The atomic configurations of each atom in these monolayer TMXCs are the same as those of conventional MXenes, but the oxidation states of the M-site atoms can be regulated by both X-site atoms and intercalated cations. This has significant implications for applications in electrochemical energy storage and surface catalysis.**

The MAX phases, with the chemical formula $M_{n+1}AX_n$, represent a large class of non-van der Waals layered ternary materials. In these compounds, M stands for an early transition metal, A is a main group or subgroup element, and X comprises carbon, nitrogen, and/or boron. The value of n is typically 1 or 2. From the perspective of layered crystal structure and chemical bonding, MAX phases can be viewed as alternating stacks of M-X and M-A sublayers along the c-axis. In the M-X sublayer, strong covalent bonds form between transition metals and nonmetal elements, while the M-A sublayer features weak metallic bonds between transition metals and main group elements such as Al and Si. By exploiting the different chemical reactivities of the two sublayers, selective chemical etching can be conducted. For example, metal A atoms in the M-A sublayer will easily lose electrons and be preferentially etched away in a proton-containing solution or Lewis acid molten salts, while the chemically-inert covalent M-X sublayer keep intact and preserve its two-dimensional structure. This selective sublayer etching process results in the creation of diverse two-dimensional *vdW* early transition metal carbides and nitrides (MXenes).

In MAX phases, not only do the [M-X] sublayers consist of covalent $[M_6X]$ octahedron, but the p orbitals of A-site chalcogens can also exhibit strong coupling with the d orbitals of M-site elements due to the high electronegativity of chalcogens, as observed in bonds such as Ti-S, Zr-S, Zr-Se, Hf-S, Hf-Se, and Hf-Te within the phases $Ti_2SC$, $Zr_2SC$, $Hf_2SC$, $Nb_2SC$, $Zr_2SB$, $Hf_2SB$, $Nb_2SB$, $Zr_2SeC$, $Hf_2SeC$, $Zr_2SeB$, $Hf_2SeB$, and $Hf_2TeB$[18-22]. The disparity between M-A bonds and M-X bonds is significantly smaller than that typically found in MAX phases, which contain both metallic and covalent bonds. Consequently, LAMS is ineffective against stripping such stable covalent bonds. In this work, the reactive difference between covalent [M-A] sublayers and covalent [M-X] sublayers is exploited and recognized. A covalent chemical scissor is introduced to selectively edit the covalent [M-B] sublayers within boride MAX phases. The customized X elements of MAX phases are orderly substituted from B, Se, S, P, to C. The resulting transition metal Xide-chalcogenides (TMXC) with unconventional X elements, derived from the covalent-sublayer edited MAX phase precursors, facilitate the discovery of a series of nanolaminated early transition metal compounds, thereby significantly broadening the spectrum of available 2D materials.

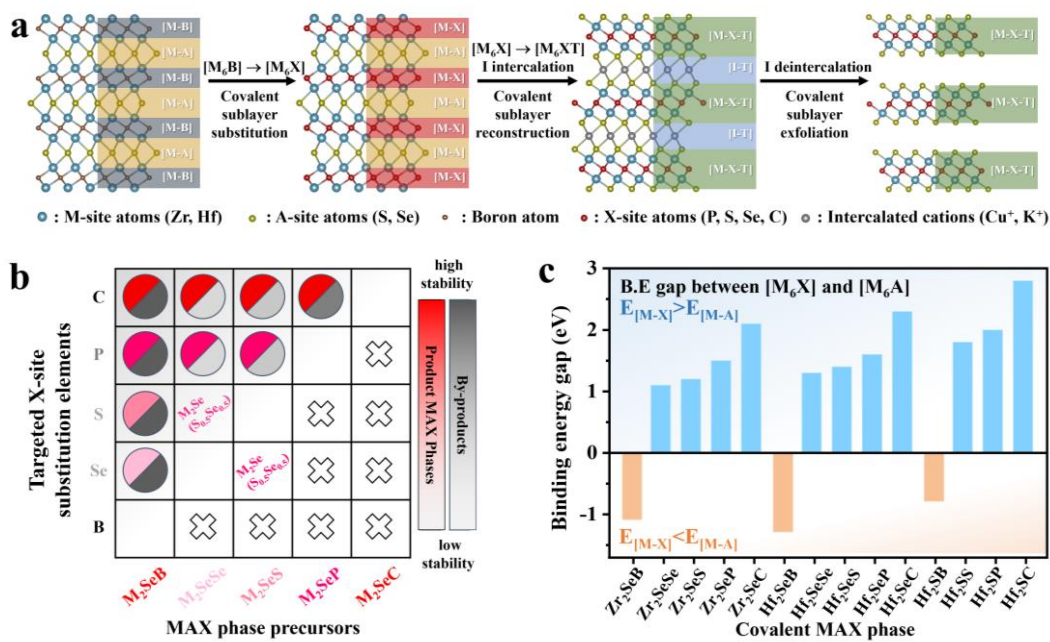

**Fig. 1.** Structural editing policy of covalent MAX phases and *i*-TMXC. (**a**) Schematic representation of the transformation from the covalent boride MAX phases to 2D TMXC nanosheets. (**b**) An illustration depicting the concept of substitution in covalent [M-X] sublayers within MAX phases. (**c**)The variations in the binding state of the M element within [$M_6X$] and [$M_6A$] configurations of the covalent MAX phase, as analyzed through XPS, whose X site represents B, Se, S, P, and C.

Fig. 1 shows the topochemical transformation from the covalent boride MAX phases to TMXC nanosheets with three sequential reaction steps: (i) the substitution of boron in the [M-B] sublayers of boride MAX phases with non-metallic atoms, facilitated by covalent chemical scissors, which reduces the Gibbs free energy due to the formation of more stable M-X covalent bonds within the MAX phases and/or the generation of ultra-stable byproducts; (ii) the intercalation of cations into the MAX phases precursor, resulting in the reconstruction of covalent [M-X-T] sublayers in the ion intercalated TMXC (*i*-TMXC); (iii) the exfoliation of the covalent [M-X-T] sublayers within *i*-TMXC through electron injection using metallic chemical scissors in an alkali molten salt environment, and the further delamination of the accordion-like alkali metal intercalated TMXC through the progressive oxidation of protons and sonication, ultimately yielding 2D TMXCs (*d*-TMXC).

The stability of transition metal compounds, which is predominantly influenced by the characteristics of chemical bonding, elucidates the transformation among boride MAX phases, selenide MAX phases, sulfide MAX phases, phosphide MAX phases, and carbide MAX phases (Fig. 1b and S1). It is noteworthy that unconventional elements such as Se, S, and P have been incorporated into the X site of MAX phases, alongside the more prevalent C and B. Furthermore, the differing valence states of metallic elements when interacting with non-metallic elements in the [M-X] and [M-A] sublayers can be leveraged to facilitate the intercalation of cations into covalent MAX phase, thus resulting in the reconstruction of the [M-X-T] sublayers (Fig. 1c). Additionally, dual anionic early transition metal compounds have been intercalated with Cu, Li, Na, and K atoms. Detailed experimental data and findings that substantiate these phases are presented in Tables S1-S4.

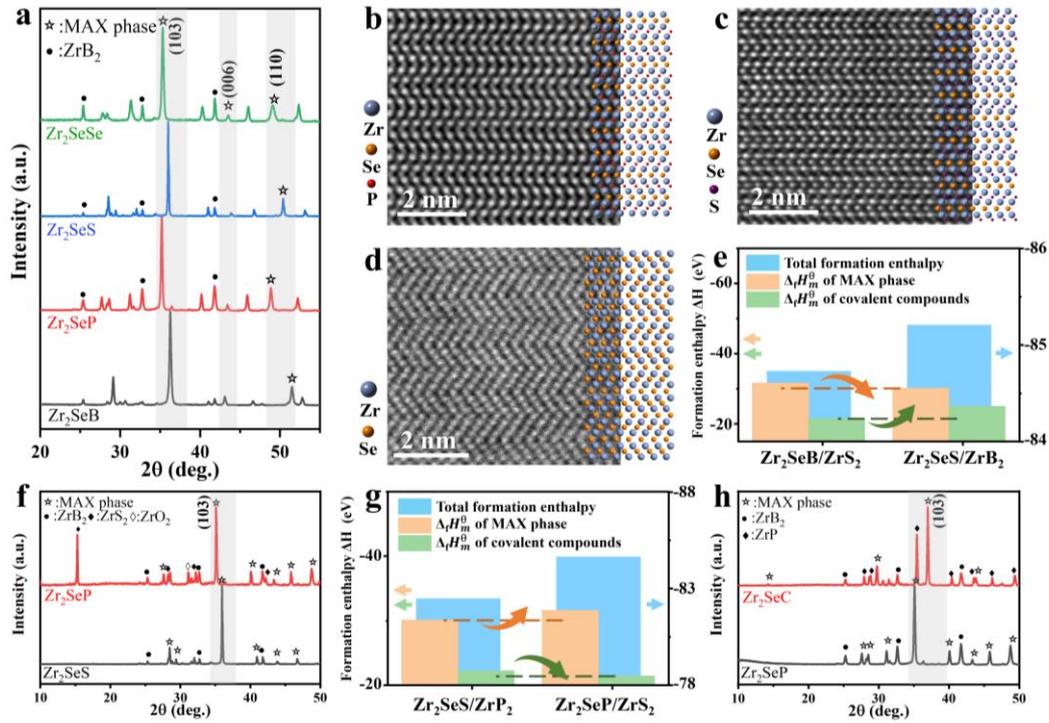

**Fig. 2.** Topochemical transformations among MAX phases utilizing covalent chemical scissors. (**a**) XRD patterns of $Zr_2SeP$, $Zr_2SeS$ and $Zr_2SeSe$ and their parent phase $Zr_2SeB$. (**b-d**) STEM images of $Zr_2SeP$, $Zr_2SeS$, and $Zr_2SeSe$ along the $[11\bar{2}0]$ zone axis with their atomic structural models, respectively. (**e**) Formation enthalpy of reactants and products involved in the transformation of $Zr_2SeB$ to $Zr_2SeS$. (f) XRD patterns of $Zr_2SeP$ and its parent phase $Zr_2SeS$. (**g**) Formation enthalpy of reactants and products involved in the transformation of $Zr_2SeS$ to $Zr_2SeP$. (**h**) XRD patterns of $Zr_2SeC$ and its parent phase $Zr_2SeP$.

Substitution of Covalent [M-X] sublayers within MAX phases is mediated by covalent chemical scissors. The covalent chemical scissors, $ZrP_2$, inducing molten state (Table S5) at reaction temperature can promote the diffusion of activated ions through a solid-state matrix, as shown in Eqs. S1-S3 (Figs. 2a-b and S2). The ionized $P^{2-}$ oxidizes the Zr atoms within $Zr_2SeB$ (Figs. S2d, e and S3c). New $Zr_2SeP$ MAX phase consequently forms and releases $B^{2-}$. Then ultra-stable $ZrB_2$ rapidly nucleates as soon as $B^{2-}$ diffuses out from the MAX phase matrix and interacts with surrounding $Zr^{4+}$. This precipitation of $ZrB_2$ lowers the Gibbs free energy of the reaction system (Fig. 2e). Similarly, Se and S can selectively substitute B within $Hf_2SB$, $Zr_2SeB$, and $Hf_2SeB$ using the corresponding chalcogenide chemical scissors, $ZrSe_2$, $HfS_2$, and

HfSe$_2$. A variety of MAX phases including Zr$_2$SeP, Hf$_2$SP, Hf$_2$SeP, Zr$_2$SeS, and Hf$_2$SeS have been successfully synthesized, as shown in Eqs. 1 and S4-S7 (Figs. S2, and S4-S7). The crystal structures of these newly formed MAX phases were thoroughly characterized using X-ray diffraction (XRD) aided with Rietveld refinement, scanning transmission electron microscopy (STEM), energy dispersive spectroscopy (EDS), and X-ray photoelectron spectroscopy (XPS). In the XRD pattern of Zr$_2$SeP, the (110) peak shifts to a lower Bragg angle as compared with Zr$_2$SeB precursor (Fig. 2a), indicating an expansion of the lattice parameter $a$ from 3.561 Å to 3.735 Å. This anticipated increase is attributed to the enlargement of the [Zr-X] sublayers, where the smaller B (97 pm) is replaced by the larger P (109 pm). The distortion of the [Zr$_6$X] octahedron is reduced due to the smaller size difference (SD) between the stacked atoms (SD$_{Zr-B}$=61 pm, and SD$_{Zr-P}$=49 pm). Notably, the changes in the parameter $c$ (0.5~0.8%) are significantly less pronounced than those observed for $a$ (3~5%) when separately comparing Zr$_2$SeP, Hf$_2$SeP, and Hf$_2$SP to Zr$_2$SeB, Hf$_2$SeB, and Hf$_2$SB (Table S2). Consequently, the M-A bond length influenced by the expanded $a$ and relatively invariable $c$ elongates during the topochemical transformation, leading to a weakening of the M-A bond comprised of identical atoms (Fig. S1). Despite the increased formation enthalpy of the edited MAX phase, the stability of the metal boride and the resultant reduction in the total Gibbs free energy of the products facilitate the topochemical transformation (Fig. 2e). The typical mirrored zig-zag arrangements of the MAX phases were observed using the atom-resolved high-angle annular dark-field (HAADF) technique in STEM (Fig. 2b-c). EDS and XPS analyses further confirmed the incorporation of substituted non-metallic elements in the as-edited MAX phase (Figs. S2-S7).

$$2Zr_2SeB + ZrP_2 = 2Zr_2SeP + ZrB_2 \qquad (1)$$

MAX phases are usually considered as ternary compounds due to the constraints associated with the X-site constituents (usually C, N, B)[22]. While Se has been identified as a suitable X-site occupant in the $Zr_2Se(B_{1-x}Se_x)$, the thermodynamically stable binary compound $Zr_2SeSe$ has not been successfully synthesized using the same powder metallurgical techniques applied to $Zr_2Se(B_{1-x}Se_x)$[23, 24]. In this study, we utilized covalent chemical scissors, specifically $ZrSe_2$, which shares the same anion as the A-site element in the $Zr_2SeB$ precursor (Fig. S8). The inherent twinned [$Zr_6Se$] triangular prism sublayers provide an architecture for the formation of $Zr_2SeSe$, thereby mitigating the need for excessive thermal energy during nucleation[1]. Furthermore, competing phases such as $Zr_2Se_3$ and $Zr_3Se_2$ can be effectively eliminated through a straightforward topochemical transformation[23] (Eq. S8). Accordingly, other binary MAX phases, including $Hf_2SS$ and $Hf_2SeSe$, were also achieved using $HfS_2$ and $HfSe_2$ as chemical scissors to tailor the $Hf_2SB$ and $Hf_2SeB$ precursors, respectively (Eqs. S9-S10) (Figs. S9-S10). In binary MAX phases, the [$M_6X$] octahedron exhibits a more regular configuration, and the alleviation of octahedral distortion leads to reduced discrepancies in atomic arrangements and conduction behavior along the *a* and *c* axes (Fig. 2d)[23].

In addition to the [M-X] sublayers derived from boride MAX phases, the new covalent [M-X] sublayers can be further substituted through the application of covalent chemical scissors, following the sequence of Se, S, P, and C. For example, $Zr_2SeC$ can be transformed from $Zr_2SeB$ precursor by sequentially utilizing $ZrS_2$, $ZrP_2$ and $ZrC$, as illustrated in Eqs. S4 and S11-S12 (Figs. 1b, 2f-h, and S11). In these reactions, the lower formation enthalpy ($Zr_2SeSe≈Zr_2SeS>Zr_2SeP>Zr_2SeC$) determined by the stronger MX bonds within as-edited MAX phase plays a predominant role (Table S4). The resultant thermodynamically stable MAX phase, following this editing process, is deemed suitable for high-temperature applications. However, the substitution process involving S and Se does not demonstrate a distinct substitution relationship. This is attributed to the comparable binding energies and valence states of these two elements when they occupy the X site in conjunction with the metal M. Instead, the decrease in free energy influenced by entropy changes, facilitates the mixing of S and Se at the X site, leading to the formation of $Hf_2Se(S_{0.5}Se_{0.5})$ (Fig. S12). Furthermore, a multi-principal MAX phase, $Hf_2Se(S_{0.25}Se_{0.25}P_{0.25}B_{0.25})$, has been successfully synthesized (Fig. S13). The grain size (~submicron) is quite small because of the refinement by the precipitation of $HfB_2$ which impedes the migration of grain boundaries within the MAX phase (Figs. S3a, d, g and S13b). Additionally, chalcogens exhibit a preferential occupancy at the A site when they coexist during the reaction. The anticipated $Hf_2SSe$ could not be synthesized using $HfSe_2$ as chemical scissors to tailor the $Hf_2SB$ precursor; instead, $Hf_2SeS$ was successfully obtained, analogous to the products derived from the $Hf_2SeB-HfS_2$ reactants, as shown in Eqs. S13-S14 (Fig. S14). Lattice parameters calculated via first-principles methods indicate a significant distortion in the $Hf_2SSe$ lattice as compared with $Hf_2SeS$. Consequently, $Hf_2SSe$ spontaneously converts to $Hf_2SeS$ due to the considerably lower in-plane energy of the latter under the similar total energy (Table S4). This observation suggests that the covalent chemical scissors can effectively customize not only the [M-X] sublayers but also the [M-A] sublayers within covalent MAX phases.

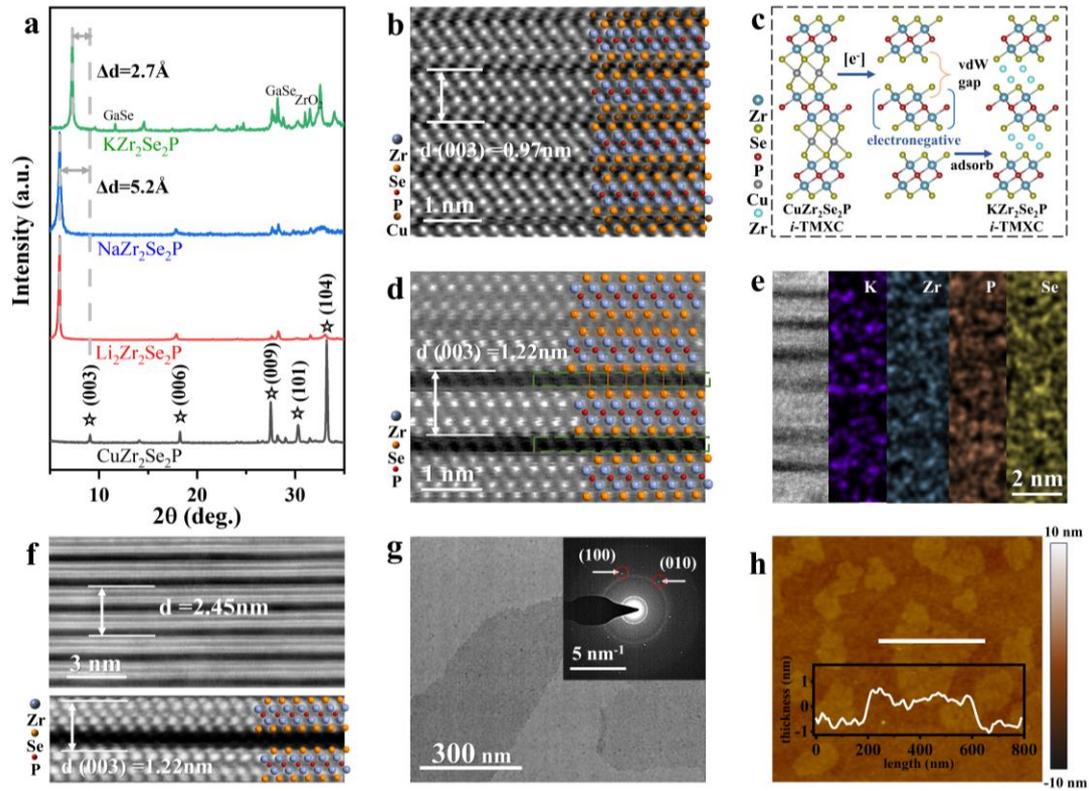

**Fig. 3.** The intercalation and deintercalation of $Zr_2Se_2P$. (**a**) XRD patterns of alkali metals intercalated $Zr_2Se_2P$ and their parent phase $CuZr_2Se_2P$. (**b**) STEM image showing atomic arrangements of $CuZr_2Se_2P$ along [110] direction, respectively. (**c**) Gap-opening mechanism of *i*-TMXC through Ga scissor. (**d-e**) STEM image along [110] direction with EDS mapping of $KZr_2Se_2P$. (**f**) STEM image showing atomic arrangements of $Zr_2Se_2P$. (**g**) TEM images of $Zr_2Se_2P$ single sheet. The inset shows the corresponding selected area electron diffraction pattern. (**h**) Atomic force microscope image of $Zr_2Se_2P$ single sheet.

Topotactic structural transformation from MAX phase to TMXC is aided by intercalation and deintercalation. The covalent MAX phases, upon X-site modification, significantly influence the composition of [M-X] sublayers[25]. This alteration serves as a foundation for the synthesis of MXene-like materials, which exhibit distinctive properties linked to M-X orbital interactions. Unfortunately, the delamination of covalent MAX phases into 2D materials is infrequently achieved. The vdW multilayered transition metal carbo-chalcogenides (TMCC) can be obtained from carbide MAX phases through a sequential process involving S oxidation and Cu intercalation; their delamination is challenging due to limited interlamellar spacing[26].

*Michael Naguib* et al introduced an electrochemical lithiation technique to facilitate the swelling of multilayered $Nb_2S_2C$ and $Ta_2S_2C$, enabling their delamination into individual sheets[27]. In this work, LAMS is employed to intercalate foreign Cu and chalcogen into the covalent MAX phase precursors, resulting in formation of *i*-TMXC. The multilayered *i*-TMXC is subsequently produced through the reduction of metal scissors in alkali molten salt, which concurrently expands the interlamellar spacing of TMXC. The TMXC nanosheets (*d*-TMXC) are then exfoliated through progressive oxidation of protons and physical delamination. For example, the topochemical transformation from $Zr_2SeP$ to 2D $Zr_2Se_2P$ is shown in Eqs. 2-4:

$$Zr_2SeP + CuSe = CuZr_2Se_2P \qquad (2)$$

$$2CuZr_2Se_2P + Ga + 2KCl = 2KZr_2Se_2P + GaCl_2\uparrow + Cu \qquad (3)$$

$$2KZr_2Se_2P + 2H^+ = 2K^+ + 2Zr_2Se_2P + H_2\uparrow \qquad (4)$$

The ionized $Cu^{2+}$ derived from LAMS scissors $CuCl_2$ exhibits a significant electron affinity and oxidizes the Zr within $Zr_2SeP$ (Eqs. S15-S16). The resultant positively charged $Zr_2SeP^+$ engages in Coulombic interactions with the surrounding $Se^{2-}$ anion, leading to the formation of $Zr_2SeP^-$ (Eq. S17). The reduced $Cu^+$ intercalated into the solid matrix is contributed to the electrical neutrality of $CuZr_2Se_2P$ (Eq. S18). Valence state analyses of Zr and Cu in $Zr_2SeP$, $CuZr_2Se_2P$, and CuSe, conducted through XPS and XFAS, elucidate the electron transfer mechanisms involved (Figs. S4d and S15e-g). Given that Zr is coordinated to both Se and P in the original MAX phase, the intercalation sites for the foreign Cu and Se can be categorized into two configurations: $[Zr_6Se]$ and $[Zr_6P]$. However, the lower valence state of Zr in the $[Zr_6Se]$, coupled with the comparatively weaker bond strength of Zr-Se relative to Zr-P interactions in the $[Zr_6P]$, suggests a preference for intercalation at the $[Zr_6Se]$ and reconstruct covalent [Zr-P-Se] sublayers (Fig. S15g and Table S4).

This preference is experimentally validated by the observation of alternating stacked new [Se$_6$Cu] octahedral sublayers alongside the native [Zr$_6$P] octahedral sublayers in STEM (Fig. 3b). The molar ratios of Cu:Zr:Se:P and the microscopic layered morphology further substantiate the topochemical transformation from Zr$_2$SeP to CuZr$_2$Se$_2$P (Fig. S15a-b). In addition, Zr$_2$SeP undergoes decomposition into ZrP and ZrSe$_2$ through the direct oxidation of Se; conversely, Zr$_2$SeTeP can be synthesized via the oxidation of Zr$_2$SeP by Te without the necessity of Cu (Fig. S16). Cu acts as an electron donor, mitigating the excessive oxidation of Zr by Se, which is characterized by high electronegativity. This phenomenon parallels the transformation observed from Nb$_2$SC to Nb$_2$S$_2$C, where Nb$_2$SC decomposes into NbC and NbS$_2$ when S serves as the oxidant[26]. Therefore, the selection of an appropriate oxidant with a weaker electron affinity than the A-site element is critical for the successful intercalation.

In order to exfoliate the TMXC nanosheets, the interlamellar spacing of CuZr$_2$Se$_2$P is opened through the reduction of Ga metal scissors in conjunction with the concurrent intercalation of K$^+$ in the molten salt. This process is analogous to the use of metal scissors for the removal of terminations from MXenes and the intercalation of metals in transition metal dichalcogenides (TMDC)[12, 27]. The electron donated from Ga reduces the Cu$^+$ within CuZr$_2$Se$_2$P and Zr$_2$Se$_2$P$^-$ consequently forms (Eqs. S19-S20). The surrounding K$^+$ subsequently intercalates into the position previously occupied by Cu$^+$ (Eqs. S21-S22) (Fig. 3c). Furthermore, the generation of low boiling-point GaCl$_2$ promotes the topochemical reaction and expands the interlamellar spacing (Eq. S23). The leftward shift of the (003) XRD peak indicates a successful topotactic structural transformation from CuZr$_2$Se$_2$P to KZr$_2$Se$_2$P (Fig. 3a). The SEM image of the fully etched CuZr$_2$Se$_2$P exhibits a characteristic accordion-like morphology, and EDS data confirms the complete removal of interlayer Cu and the successful intercalation of K (Fig. S17a-b). The STEM image of K-intercalated $Zr_2Se_2P$ demonstrates a significant increase in interlayer spacing, with the interplanar distance of (003) expanding from 0.97 nm to 1.22 nm. This observation aligns with the XRD results, and EDS mapping further corroborates the intercalation of K between the $Zr_2Se_2P$ single layers (Fig. 3d, e). This methodology is broadly applicable for the intercalation of other alkali metals, such as Li and Na (Eqs. S24-S25). The resulting Li or Na intercalated TMXCs exhibit a greater swelling of (003) (5.2 Å) as compared with that of $KZr_2Se_2P$ (2.7 Å) (Figs. 3a and S18). Notably, XFAS analysis indicates that the valence state of Zr in Li intercalated $Zr_2Se_2P$ is equivalent to that in $Zr_2Se_2P$, but lower than in $CuZr_2Se_2P$, $KZr_2Se_2P$ and $NaZr_2Se_2P$ (Fig. S18f). This suggests that a greater amount of $Li^+$ ions can be intercalated into the $Zr_2Se_2P$ matrix as compared with the latter three. This finding implies that diverse intercalated TMXCs can be achieved through the mediation of molten salts with varying coordination environments of metallic cations[28, 29]. The $Zr_2Se_2P$ nanosheets were obtained from the oxidation of $KZr_2Se_2P$ using protons and physical delamination. The reaction of $H^+$ with $KZr_2Se_2P$ generates hydrogen gas ($H_2$), and facilitates the removal of alkali metal ions from the vdW gap of the $Zr_2Se_2P$ layers (Fig. 3f). This interlayer hydrogen evolution induce further swelling of multilayered $Zr_2Se_2P$, ultimately leading to the delamination process into $Zr_2Se_2P$ nanosheets (~monolayer) (Fig. 3g-h)[30]. Unlike the typical MXenes derived from the exfoliation of metallic A-site elements within MAX phases, which are characterized by a single selenium termination[31], 2D $Zr_2Se_2P$ exhibits double selenium terminations on its surface, indicating a higher oxidation state of Zr. Consequently, these 2D TMXCs, which involve previous TMCC, can be classified as oxidized MXenes. This approach enables the creation of diverse 2D TMXCs with dual anions (XC). The modification of orbital coupling within the [M-X] octahedral sublayers prompted by the terminations, is anticipated to yield intriguing properties in 2D materials. For example, the various interaction modes between alkali metals and Se regulated by [M-X] sublayers, suggesting that TMXC may be employed as a cathode material for alkali metal/Se batteries[32]. The band gap exceeding 1.5 eV found in the $Li_2Zr_2Se_2P$ indicates its application as optoelectronic devices (Fig. S19).

**Conclusion:** In summary, we have developed a methodology that employs chemical scissors to edit the covalent sublayers of nanolaminated early transition metal compounds, specifically MAX phases and *i*-TMXCs. The composition of the covalent [M-X] sublayers within MAX phases is precisely regulated through the covalent chemical scissors. Additionally, unconventional elements (such as Se, S, and P) are substituted into the X site of MAX phases via topochemical transformation, a process that is not feasible with traditional powder metallurgy techniques. The resulting 2D TMXCs, which incorporate two anions and are derived from the tailored covalent MAX phases successively using LAMS and metal scissors, exhibit unique structural characteristics akin to both MXenes and transition metal chalcogenides. This advancement significantly expands the family of 2D materials. Considering the diverse coordination environments present within TMXCs, further investigations may reveal intriguing physical properties, which could lead to progress in fields ranging from optoelectronics to energy storage.